\documentclass[a4paper,11pt]{PoS}

\usepackage{amsmath}
\usepackage{subfigure}

\newcommand{\be}{\begin{equation}}
\newcommand{\ee}{\end{equation}}
\newcommand{\bea}{\begin{eqnarray}}
\newcommand{\eea}{\end{eqnarray}}

\title{Electromagnetic corrections to the leptonic decay rates of charged pseudoscalar mesons: lattice results}

\ShortTitle{Electromagnetic corrections to the leptonic decay rates ...}

\author{V.~Lubicz$^{(a,b)}$, G.~Martinelli$^{(c)}$, C.T.~Sachrajda$^{(d)}$, F.~Sanfilippo$^{(d)}$, \speaker{S.~Simula}$^{(b)}$, N.~Tantalo$^{(e)}$, C.~Tarantino$^{(a,b)}$

\\

\it $^{(a)}$ Dipartimento di Matematica e Fisica, Universit\`a  Roma Tre, Rome, Italy.\\ Email: \email{lubicz@fis.uniroma3.it}

\it $^{(b)}$ Istituto Nazionale di Fisica Nucleare, Sezione di Roma Tre, Rome, Italy.\\ Email: \email{simula@roma3.infn.it}

\it $^{(c)}$ Dipartimento di Fisica, Universit\`a  La Sapienza and INFN, Sezione di Roma, Rome, Italy.\\ Email: \email{martinelli@roma1.infn.it}

\it $^{(d)}$ School of Physics \& Astronomy, University of Southampton, Southampton, United Kingdom.\\ Email: \email{cts@soton.ac.uk}

\it $^{(e)}$ Dipartimento di Fisica, Universit\`a  di Roma and INFN Tor Vergata, Rome, Italy.\\ Email: \email{tantalo@roma2.infn.it}

}

\abstract{Electromagnetic effects in the leptonic decay rates $\pi^+ \to \mu^+ \nu$ and $K^+ \to \mu^+ \nu$ are evaluated for the first time on the lattice. Following a method recently proposed in Ref.~\cite{Carrasco:2015xwa} the emission of virtual photons at leading order in the electromagnetic coupling is evaluated on the lattice and the infrared divergence computed for a point-like meson at finite lattice volume is subtracted. The physical decay rate is then obtained by adding the emission of real and virtual photons regularised with a photon mass. Using the gauge ensembles produced by the European Twisted Mass Collaboration with $N_f = 2 + 1 + 1$ dynamical quarks the feasibility of our approach is demonstrated. Preliminary results for the electromagnetic corrections to charged (neutral) pion and kaon masses as well as to the leptonic decay rates of charged pions and kaons are presented.}

\FullConference{34th annual International Symposium on Lattice Field Theory\\
		24-30 July 2016\\
		University of Southampton, UK}

\begin{document}

\section{Introduction}
\label{sec:intro}

The determination of several hadronic quantities relevant for flavour physics phenomenology using large-scale QCD simulations on the lattice has reached an impressive degree of precision \cite{FLAG} such that both electromagnetic (e.m.) and strong isospin-breaking (IB) effects cannot be neglected anymore.
In the past few years, using different methodologies, accurate lattice results including e.m.~effects have been obtained for the hadron spectrum, as in the case of the neutral-charged mass splittings of light pseudoscalar (PS) mesons and baryons (see, e.g., Refs.~\cite{deDivitiis:2013xla,Borsanyi:2014jba}).
However, while in the calculation of e.m.~effects in the hadron spectrum no infrared (IR) divergencies can appear, in the case of other hadronic quantities, like the decay amplitudes, IR divergencies are present and can be cancelled out in the physical observable only by summing up diagrams containing both real and virtual photons \cite{BN37}.
This is the case, for example, of the leptonic $\pi_{\ell 2}$ and $K_{\ell 2}$ and of the semileptonic $K_{\ell 3}$ decay rates, which play a crucial role for an accurate determination of the CKM entries $|V_{us}|$ and $|V_{us} / V_{ud}|$. 
We stress that in the case of inclusive decay rates it is not enough to add the electromagnetic interaction to the quark action, because amplitudes with different numbers of real photons must be evaluated separately, before being combined in the inclusive rate. 

The occurrence of IR divergencies requires therefore the use of a new strategy different from those developed to compute the e.m.~effects in the hadron spectrum.
Such a new strategy was proposed in Ref.~\cite{Carrasco:2015xwa}, where the determination of the inclusive decay rate of a charged pseudoscalar (PS) meson into either a final $\ell^\pm \bar{\nu}_\ell$ pair or a final $\ell^\pm \bar{\nu}_\ell \gamma$ state was addressed.
A crucial condition for the strategy of Ref.~\cite{Carrasco:2015xwa} is that the maximum energy $\Delta E_\gamma$ of the emitted photon (in the PS-meson rest frame) has to be small enough in order not to resolve the internal structure of the decaying PS meson. 
In this way the inclusive rate $\Gamma(PS^+ \to \ell^+ \bar{\nu}_\ell [\gamma])$ can be expressed as \cite{Carrasco:2015xwa}
 \be
     \Gamma(PS^\pm \to \ell^\pm \bar{\nu}_\ell [\gamma]) = \mbox{lim}_{L \to \infty} [\Gamma_0(L) -  \Gamma_0^{pt}(L)] + 
     \mbox{lim}_{\mu_\gamma \to 0} [\Gamma_0^{pt}(\mu_\gamma) + \Gamma_1^{pt}(\Delta E_\gamma, \mu_\gamma)] ~ ,
     \label{eq:Gamma}
 \ee
where the subscripts $0$ and $1$ indicate the number of photons in the final state, while the superscript ``pt'' denotes the point-like treatment of the decaying PS meson.
In the r.h.s.~of Eq.~(\ref{eq:Gamma}) the terms $\Gamma_0(L)$ and $\Gamma_0^{pt}(L)$ can be evaluated on the lattice using the lattice size $L$ as an IR regulator.
Their difference is free from IR divergencies and therefore the limit $L \to \infty$ can be performed obtaining a result independent on the specific IR regularization \cite{Carrasco:2015xwa}.
In a similar way the terms $\Gamma_0^{pt}(\mu_\gamma)$ and $\Gamma_1^{pt}(\Delta E\gamma, \mu_\gamma)$ can be calculated perturbatively using a photon mass $\mu_\gamma$ as an IR regulator.
Their sum is free from IR divergencies thanks to the Bloch-Nordsiek mechanism \cite{BN37}, so that the limit $\mu_\gamma \to 0$ can be performed obtaining again a result independent on the specific IR regularization.

The explicit calculation of $\mbox{lim}_{\mu_\gamma \to 0} [\Gamma_0^{pt}(\mu_\gamma) + \Gamma_1^{pt}(\Delta E\gamma, \mu_\gamma)]$ has been carried out in Ref.~\cite{Carrasco:2015xwa}, while the calculation of $\Gamma_0^{pt}(L)$ has been presented separately at this conference \cite{Tantalo}.

In this contribution we describe the first lattice calculation of the QCD+QED effects in $\Gamma_0(L)$ and we present preliminary results for the inclusive decay rates of charged pions and kaons.

\section{Simulation details}
\label{sec:simulations}

The gauge ensembles used in this contribution are the ones generated by the European Twisted Mass Collaboration (ETMC) with $N_f = 2 + 1 + 1$ dynamical quarks, and the lattice actions for sea and valence quarks are the same used in Ref.~\cite{Carrasco:2014cwa} to determine the up, down, strange and charm quark masses.
We considered three values of the inverse bare lattice coupling $\beta$ and different lattice volumes, as shown in Table \ref{tab:simudetails}, where the number of configurations analysed ($N_{cfg}$) corresponds to a separation of $20$ trajectories.
At each lattice spacing, different values of the light sea quark masses have been considered. 
The light valence and sea quark masses are always taken to be degenerate. 
The bare mass of the strange valence quark ($a\mu_s$) is obtained, at each $\beta$, using the physical strange mass and the mass renormalization constants determined in Ref.~\cite{Carrasco:2014cwa}.
The values of the lattice spacing are $a = 0.0885(36)$, $0.0815(30)$, $0.0619(18)$ fm at $\beta = 1.90$, $1.95$ and $2.10$, respectively.
The value of the final lepton mass has been taken fixed at its physical value.

\begin{table}[htb!]
{\scriptsize
\begin{center}
\begin{tabular}{||c|c|c|c|c|c|c|c|c|c||}
\hline
ensemble & $\beta$ & $V / a^4$ &$a\mu_{sea}=a\mu_{val}$&$a\mu_\sigma$&$a\mu_\delta$&$N_{cfg}$& $a\mu_s$ & $M_\pi {\rm (MeV)}$ & $M_K {\rm (MeV)}$ \\
\hline \hline
$A30.32$ & $1.90$ & $32^{3}\times 64$ &$0.0030$ &$0.15$ &$0.19$ &$150$&  $0.02363$ & 275 & 577 \\
$A40.32$ & & & $0.0040$ & & & $100$ & & 315 & 588 \\
$A50.32$ & & & $0.0050$ & & &  $150$ & & 350 & 595 \\
\cline{1-1} \cline{3-4} \cline{7-7} \cline{9-10}
$A40.24$ & & $24^{3}\times 48 $ & $0.0040$ & & & $150$ & & 324 & 594 \\
$A60.24$ & & & $0.0060$ & & &  $150$ & & 388 & 610 \\
$A80.24$ & & & $0.0080$ & & &  $150$ & & 438 & 624 \\
$A100.24$ &  & & $0.0100$ & & &  $150$ & & 497 & 650 \\
\cline{1-1} \cline{3-4} \cline{7-7} \cline{9-10}
$A40.20$ & & $20^{3}\times 48 $ & $0.0040$ & & & $150$ & & 330 & 597 \\
\hline \hline
$B25.32$ & $1.95$ & $32^{3}\times 64$ &$0.0025$&$0.135$ &$0.170$& $150$& $0.02094$ & 259 & 553 \\
$B35.32$ & & & $0.0035$ & & & $150$ & & 300 & 562 \\
$B55.32$ & & & $0.0055$ & & & $150$ & & 377 & 587 \\
$B75.32$ &  & & $0.0075$ & & & $~80$ & & 437 & 608 \\
\cline{1-1} \cline{3-4} \cline{7-7} \cline{9-10}
$B85.24$ & & $24^{3}\times 48 $ & $0.0085$ & & & $150$ & & 463 & 617 \\
\hline \hline
$D15.48$ & $2.10$ & $48^{3}\times 96$ &$0.0015$&$0.1200$ &$0.1385 $& $100$& $0.01612$ & 224 & 538 \\ 
$D20.48$ & & & $0.0020$  &  &  & $100$ & & 255 & 541 \\
$D30.48$ & & & $0.0030$ & & & $100$ & & 310 & 554 \\
 \hline   
\end{tabular}
\end{center}
}
\vspace{-0.25cm}
\caption{\it \footnotesize Values of the simulated sea and valence quark bare masses, of the pion ($M_\pi$) and kaon ($M_K$) masses for the $16$ ETMC gauge ensembles with $N_f = 2+1+1$ dynamical quarks adopted in this contribution (see Ref.~\cite{Carrasco:2014cwa}). The values of the strange quark bare mass $a \mu_s$, given for each gauge ensemble, correspond to the physical strange quark mass $m_s = 99.6 (4.3)$ MeV determined in Ref.~\cite{Carrasco:2014cwa}.}
\label{tab:simudetails}
\end{table}

\section{Evaluation of the relevant amplitudes}
\label{sec:master}

According to Ref.~\cite{Carrasco:2015xwa} the inclusive decay rate (\ref{eq:Gamma}) can be written as
 \be
      \Gamma(PS^\pm \to \ell^\pm \bar{\nu}_\ell [\gamma]) =  \Gamma^{(tree)} \cdot \left[ 1 + \delta R_{PS} \right] ~ ,
      \label{eq:master}
 \ee
where $\Gamma^{(tree)}$ is the tree-level decay rate given by
 \be
     \Gamma^{(tree)} = \frac{G_F^2}{8 \pi} |V_{q_1 q_2}|^2 m_\ell^2 \left( 1 - \frac{m_\ell^2}{M_{PS}^2} \right)^2 \left[ f_{PS}^{(0)} \right]^2 M_{PS} ~ ,
     \label{eq:Gamma0}
 \ee
 and
 \be
    \delta R_{PS} = \alpha_{em} \frac{2}{\pi} \mbox{log}\left( \frac{M_Z}{M_W} \right) + 2 \frac{\delta A_{PS}}{A_{PS}^{(0)}} - 
                             2 \frac{\delta M_{PS}}{M_{PS}^{(0)}} + \delta \Gamma^{(pt)}(\Delta E_\gamma) ~ .
    \label{eq:RPS}
 \ee
The quantity $\delta \Gamma^{(pt)}(\Delta E_\gamma)$ can be read off from Eq.~(51) of Ref.~\cite{Carrasco:2015xwa}, while the term containing $\mbox{log}(M_Z/M_W)$ comes from short-distance electroweak corrections, $M_{PS}$ is the PS meson mass including both e.m.~and strong IB corrections, and $A_{PS}^{(0)}$ is the QCD axial amplitude
 \be
      A_{PS}^{(0)} \equiv Z_V \langle 0 | \bar{q}_2 \gamma_0 \gamma_5 q_1 | PS^{(0)} \rangle \equiv f_{PS}^{(0)} M_{PS}^{(0)}
      \label{eq:APS0}
 \ee
where $f_{PS}^{(0)}$ defines the PS meson decay constant known in pure QCD, and $Z_V$ is the renormalization constant of the axial current in our twisted-mass setup in which the $r$-parameters of the two valence quarks are chosen to be opposite ($r_1 = - r_2$).
Throughout this contribution the superscript $^{(0)}$ indicates quantities that do not contain e.m.~and strong IB corrections.

\begin{figure}[htb!]
\parbox{10.0cm}{~~ \includegraphics[scale=0.55]{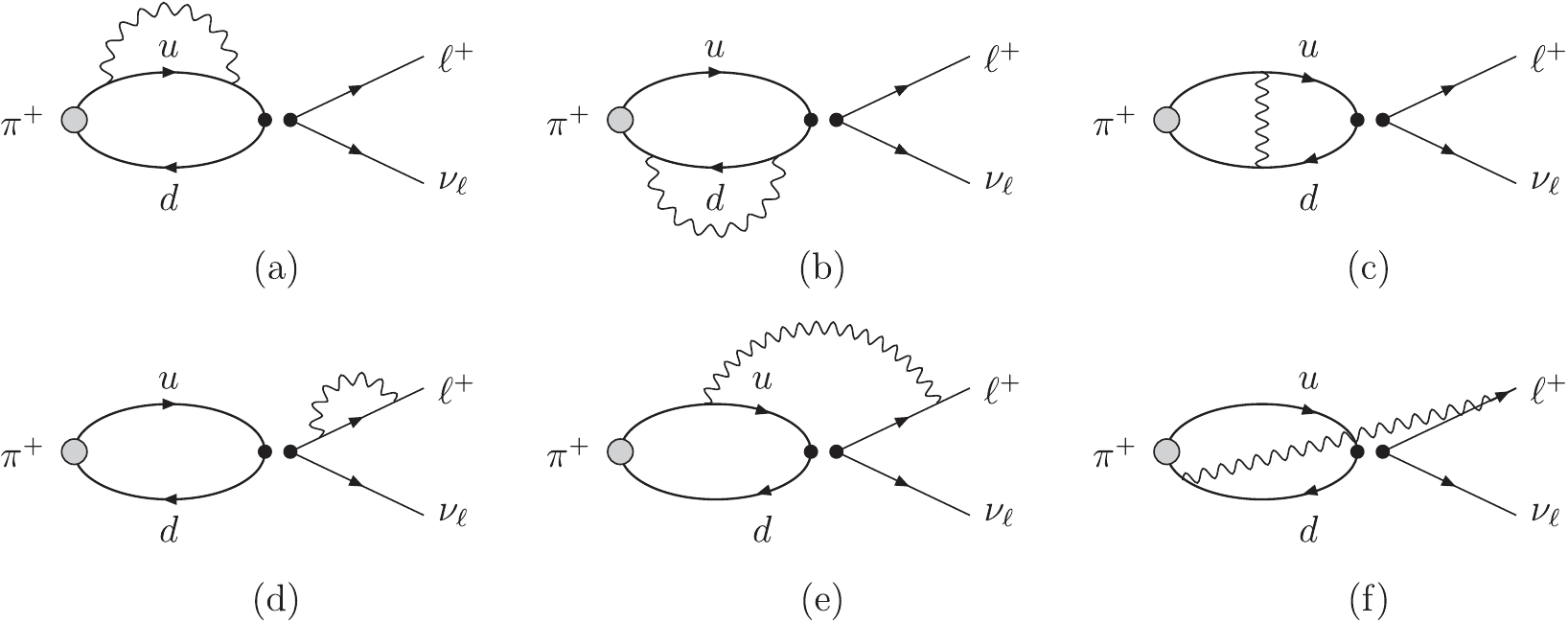}}$~$\parbox{5.0cm}{ \caption{\it \footnotesize Connected diagrams contributing at $O(\alpha_{em})$ to the amplitude for the decay $\pi^+ \to \ell^+ \nu$. The labels (a)-(f) are introduced to identify the individual diagrams when describing their evaluation in the text.}}
\label{fig:diagrams}
\end{figure}

The evaluation of $\delta A_{PS}$ and $\delta M_{PS}$ can be obtained from the diagrams shown in Figs. 5 and 6 of Ref.~\cite{Carrasco:2015xwa}.
We adopt the quenched QED approximation, which neglects the sea-quark electric charges and corresponds to consider only the connected diagrams shown in Fig.~\ref{fig:diagrams}.
In addition one should include also the contributions coming from the tadpole operator, the e.m.~shift in the critical mass and the insertion of the isovector scalar density (see Refs.~\cite{deDivitiis:2013xla,deDivitiis:2011eh}).

\begin{figure}[htb!]
\begin{center}
\includegraphics[scale=0.75]{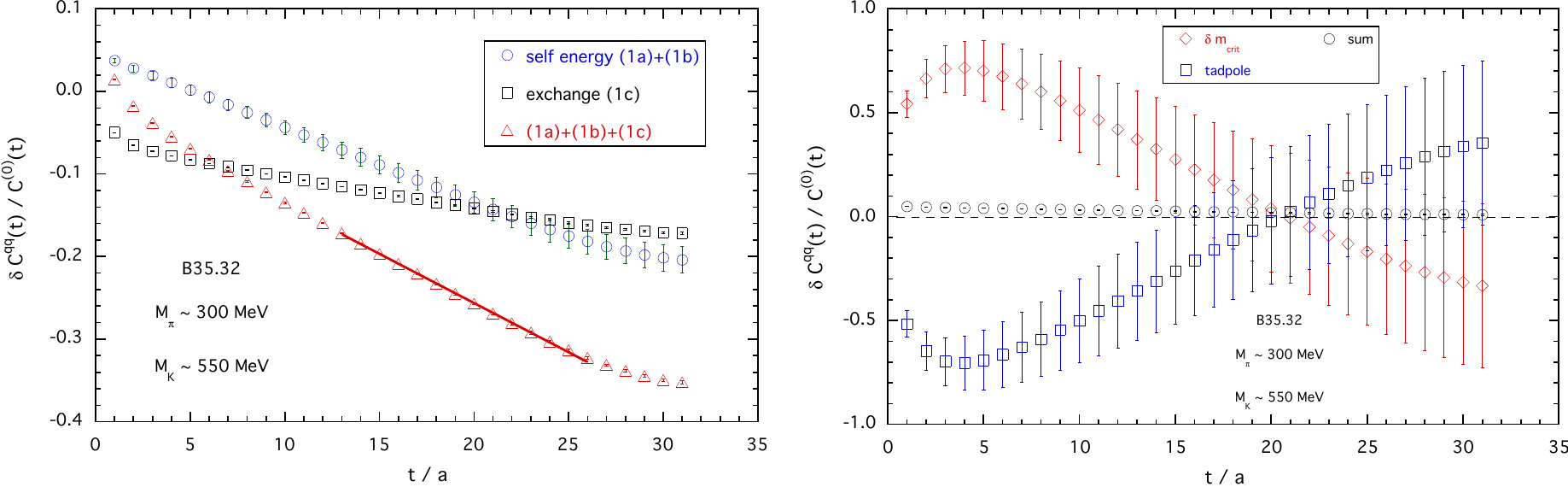}
\end{center}
\vspace{-0.70cm}
\caption{\it \footnotesize Left panel: contributions of the self-energy (\protect\ref{fig:diagrams}a)+(\protect\ref{fig:diagrams}b) and exchange (\protect\ref{fig:diagrams}c) diagrams in the case of the K-meson for the gauge ensemble B35.32. The triangles represent the sum (\protect\ref{fig:diagrams}a)+(\protect\ref{fig:diagrams}b)+(\protect\ref{fig:diagrams}c), while the solid line is the ``linear" fit (\protect\ref{eq:ratio_qq}) applied in the time interval where the ground-state is dominant. Right panel: contributions of the tadpole operator and of the e.m.~shift of the critical mass. The sum of the two terms, shown by the circles, is determined quite precisely.}
\label{fig:Cqq}
\end{figure}

The contribution $\delta A_{PS}^{qq}$ of the diagrams (\ref{fig:diagrams}a)-(\ref{fig:diagrams}c) can be extracted from the correlator
 \be
     \delta C^{qq}(t) \equiv -\frac{1}{2} \sum_{\vec{x}, x_1, x_2} \langle 0| T \left \{ J_w^\rho(0) j_\mu^{em}(x_1) j_\mu^{em}(x_2) 
                               \phi_{PS}^\dagger(\vec{x}, -t) \right \} | 0 \rangle \Delta_{em}(x_1, x_2) p_{PS}^\rho / M_{PS} ~ ,
     \label{eq:Cqq}
 \ee
where $J_w^\rho(0)$ is the quark (V-A) weak current, $j_\mu^{em}$ is the (conserved) e.m.~current\footnote{The use of the conserved e.m.~current guarantees the absence of contact terms in the product $j_\mu^{em}(x_1) j_\mu^{em}(x_2)$.}, $\phi_{PS}$ is the PS interpolating field and 
$\Delta_{em}(x_1, x_2)$ is the photon propagator.
As in Ref.~\cite{deDivitiis:2013xla} the ratio of $\delta C^{qq}(t)$ with the corresponding tree-level correlator $C^{(0)}(t) \equiv \sum_{\vec{x}} \langle 0| T \left \{ J_w^\rho(0) \phi_{PS}^\dagger(\vec{x}, -t) \right \} | 0 \rangle p_{PS}^\rho / M_{PS}$ is almost a linear function in the time distance $t$ for large values of $t$.
More precisely one gets
 \be
      \frac{\delta C^{qq}(t)}{C^{(0)}(t)} ~ _{\overrightarrow{ t >> a, (T-t) >> a }} ~ \frac{\delta [Z_{PS} A_{PS}^{qq}]}{Z_{PS}^{(0)} A_{PS}^{(0)}} + 
                                                          \frac{\delta M_{PS}}{M_{PS}^{(0)}} \left[ M_{PS}^{(0)} \left( \frac{T}{2} - t \right) \frac{e^{-M_{PS}^{(0)} t} + 
                                                          e^{-M_{PS}^{(0)} (T-t)}}{e^{-M_{PS}^{(0)} t} - e^{-M_{PS}^{(0)} (T-t)}} - 1 \right] ~ ,
      \label{eq:ratio_qq}
 \ee
where $Z_{PS} \equiv \langle 0 | \phi_{PS}(0) | PS \rangle$ is the coupling of the PS interpolating field with the ground-state.
Thus, the e.m.~correction to the PS mass, $\delta M_{PS}$, can be extracted from the slope of $\delta C^{qq}(t) / C^{(0)}(t)$, while the quantity $\delta [Z_{PS} A_{PS}^{qq}]$ from its intercept, as shown in Fig.~\ref{fig:Cqq}.

As explained in Ref.~\cite{Carrasco:2015xwa}, in order to get the quantity $\delta A_{PS}^{qq}$ the e.m.~correction $\delta Z_{PS}$ should be separately determined by evaluating a correlator similar to the one of Eq.~(\ref{eq:Cqq}), in which the weak operator $J_w^\rho p_{PS}^\rho / M_{PS}$ is replaced by the PS interpolating field $\phi_{PS}$.

We have analysed the results obtained for the e.m.~and strong IB corrections to the pion and kaon masses.
The details will be given in a forthcoming publication.
Here we anticipate that after performing the chiral extrapolation and the continuum and infinite volume limits we get the preliminary results: $M_{\pi^+}^2 - M_{\pi^0}^2 = 1.226 (58)_{stat} (96)_{syst} 10^{-3}$ GeV$^2$, $\epsilon_\gamma[\overline{MS}(2~\mbox{GeV})] = 0.833 (22)_{stat} (28)_{syst}$ and $(m_d - m_u)[\overline{MS}(2~\mbox{GeV})] = 2.69 (5)_{stat} (13)_{syst}$ MeV, where the last two systematic errors do not include the effects coming from the QED quenching approximation.

The contribution $\delta A_{PS}^{q\ell}$ of the diagrams (\ref{fig:diagrams}e)-(\ref{fig:diagrams}f) can be determined by studying the correlator\footnote{The diagram (\ref{fig:diagrams}d) (lepton wave function renormalization) can be safely omitted, since it cancels out exactly when the IR divergence evaluated for a point-like meson is subtracted.}
 \bea
     \delta C^{q\ell}(t) & \equiv & - \sum_{\vec{x}, x_1, x_2} \langle 0| T \left \{ J_w^\rho(0) j_\mu^{em}(x_1) \phi_{PS}^\dagger(\vec{x}, -t) \right \}
                                                  | 0 \rangle \Delta_{em}(x_1, x_2) e^{E_\ell t_2 - i \vec{p}_\ell \cdot \vec{x}_2} \nonumber \\
                                & \cdot & \overline{u}(p_\nu) \gamma_\rho (1 - \gamma_5) S^\ell(0, x_2) \gamma_\mu v(p_\ell) \left[ \overline{v}(p_\ell) 
                                                \gamma_\sigma (1 - \gamma_5) u(p_\nu) p_{PS}^\sigma / M_{PS} \right] ~ ,
     \label{eq:Cql}
 \eea
where $S^\ell(0, x_2)$ stands for the free twisted-mass lepton propagator, while it is understood that in the quark weak current $J_w^\rho(0)$ the vector and axial parts are renormalised according to our twisted-mass setup, namely the vector weak current renormalises with $Z_A$ and the axial one with $Z_V$.
At large time distances one has
 \be
      \delta C^{q\ell}(t) ~ _{\overrightarrow{ t >> a, (T-t) >> a }} ~ \delta A_{PS}^{q\ell} ~ Tr(p_\ell, p_{PS}) ~ Z_{PS}^{(0)}
                                                                                                    \left[ e^{-M_{PS}^{(0)} t} \pm \mbox{backward signals} \right] / 
                                                                                                    \left( 2 M_{PS}^{(0)} \right) ~ ,
      \label{eq:larget}
 \ee
where $Tr(p_\ell, p_{PS})$ stands for the tree-level leptonic trace (evaluated on the lattice using non-periodic boundary conditions), while the sign of the backward signals changes according to the spatial/temporal components of the quark (V-A) weak current. 
The backward signals can be properly subtracted by introducing the new correlator
 \be
     \delta \overline{C}^{q\ell}(t) \equiv \frac{1}{2} \left\{ \delta C^{q\ell}(t) + \left[ \delta C^{q\ell}(t-1) - \delta C^{q\ell}(t+1) \right] / 
                                                           \left( e^{M_{PS}^{(0)} } - e^{-M_{PS}^{(0)}} \right) \right\} ~ ,
    \label{eq:backward}
 \ee
so that $\delta A_{PS}^{q\ell} / A_{PS}^{(0)}$ can be extracted from the plateau of the ratio $\delta \overline{C}^{q\ell}(t) / \overline{C}^{(0)}(t)$ at large $t$.

\section{Chirality mixing and subtraction of universal finite size effects}
\label{sec:FSE}

In order to regularise the UV divergencies of the four-fermion effective theory the photon propagator has to be calculated in the so-called W-regularization scheme. 
Since the W-boson mass is too large to be simulated on the lattice, a perturbative matching for the bare lattice operator $O_1^{bare} \equiv \overline{q}_2 \gamma_\mu (1 - \gamma_5) q_1 \overline{\nu} \gamma_\mu (1 - \gamma_5) \ell$ with the W-regularization scheme has been calculated at leading order in $e^2$ in Ref.~\cite{Carrasco:2015xwa}, obtaining $O_1^{W-reg} = O_1^{bare} + \sum_{i=1}^5 Z_i O_i^{bare}$.
The operators $O_i^{bare}$ for $i = 2, ...,5$, listed in Ref.~\cite{Carrasco:2015xwa}, mix with $O_1^{bare}$ when the chiral symmetry is broken on the lattice.
We leave the details of the analysis of the effects of the chirality mixing on $\delta A_{PS}$ to a forthcoming publication.
Here we limit ourselves to quote the final result
 \be
     \delta A_{PS} = \delta A_{PS}^{qq} + \delta A_{PS}^{q\ell} + (Z_1 + Z_2) A_{PS}^{(0)} - \delta A_{PS}^{pt}(L) ~ ,
     \label{eq:deltaAPS}
 \ee
where $Z_1 + Z_2 = \alpha_{em} [5 ~ \mbox{log}(a M_W) - 8.863 + 0.536] / (4 \pi)$ (see Ref.~\cite{Carrasco:2015xwa}).

\begin{figure}[htb!]
\parbox{7.0cm}{\includegraphics[scale=0.325]{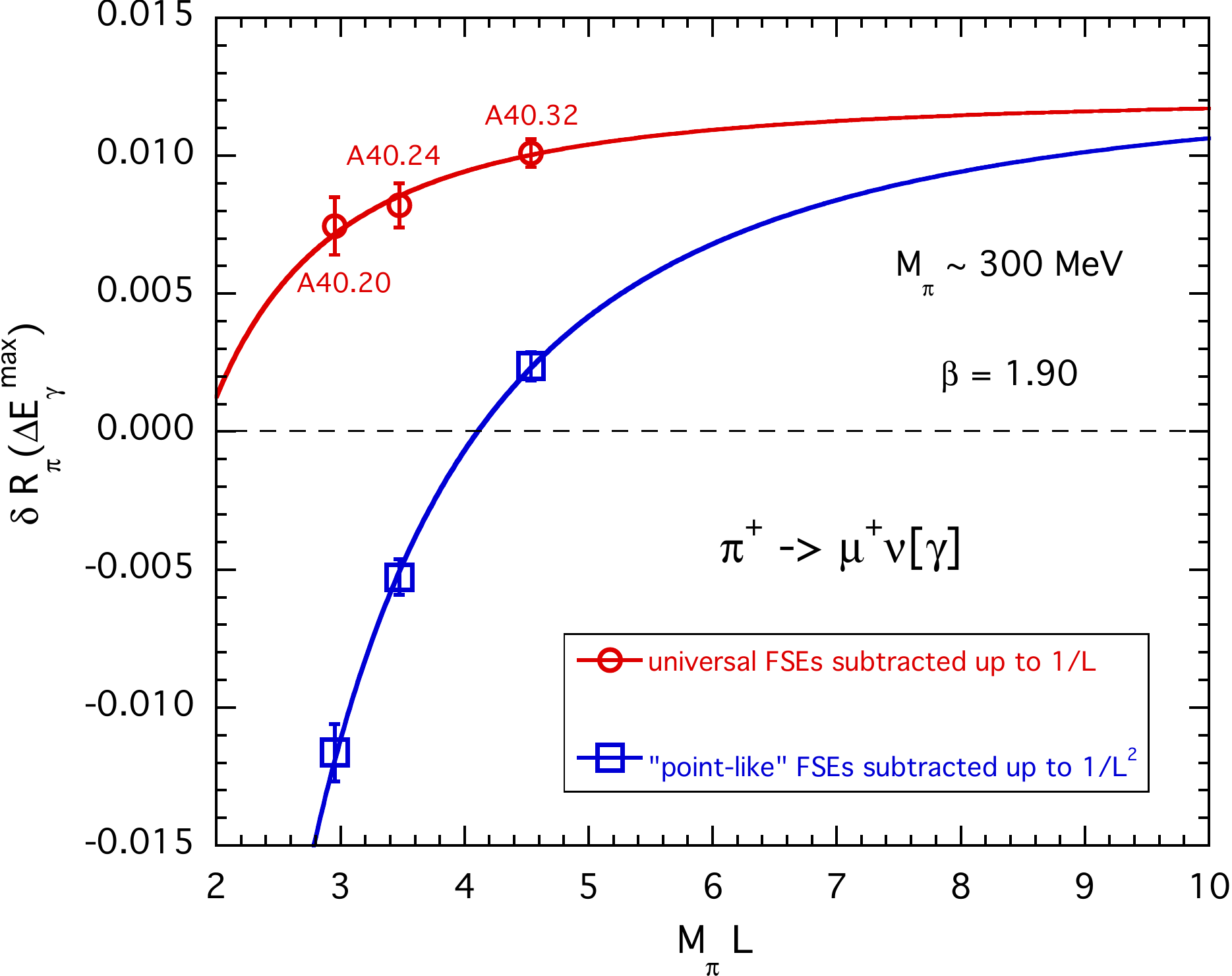}}$~$\parbox{8.0cm}{\caption{\it \footnotesize Results for the correction $\delta R_\pi(\Delta E_\gamma^{max})$ to the decay $\pi^+ \to \mu^+ \nu [\gamma]$ (see Eq.~(\protect\ref{eq:RPS})) for the gauge ensembles A40.20, A40.24 and A40.32 corresponding to the same lattice spacing ($\beta = 1.90$) and to the same pion mass ($M_\pi \sim 300$ MeV), but different lattice sizes. The red points correspond to the subtraction of the universal FSEs, i.e.~up to order $O(1/L)$ in Eq.~(\protect\ref{eq:APS_pt}), while the blue squares include also the subtraction of the ``point-like'' term $b_2 / (M_\pi L)^2$. The solid lines are the results of the simple fit $a + b / L^2$ with $a$ and $b$ being free parameters. Note that the asymptotic values of the two fits for $L \to \infty$ are in agreement. \protect\vspace{0.5cm}}}
\label{fig:FSE}
\end{figure}

In Eq.~(\ref{eq:deltaAPS}) we have included also the subtraction of the term $\delta A_{PS}^{pt}(L)$, which corresponds to the contribution of virtual photon emissions from a point-like meson using the lattice size $L$ as the IR regulator.
The quantity $\delta A_{PS}^{pt}(L)$ has been calculated in Ref.~\cite{Tantalo}, obtaining the result
 \be
     \delta A_{PS}^{pt}(L) = b_{IR} ~ \mbox{log}(M_{PS} L) + b_0 + b_1 / (M_{PS} L) + b_2 / (M_{PS} L)^2 + 
                                                                 b_3 / (M_{PS} L)^3 + O(e^{-M_{PS} L}) ~ ,
       \label{eq:APS_pt}
 \ee
where the coefficients $b_j$ ($j = IR, 0, 1, 2, 3$) depend on the mass ratio $m_\ell / M_{PS}$.
The relevant point is that structure-dependent finite size effects (FSEs) start only at order $O(1/L^2)$, i.e.~all the terms up to $O(1/L)$ in Eq.~(\ref{eq:APS_pt}) are ``universal'' \cite{Tantalo}.
The FSE subtraction (\ref{eq:APS_pt}) is illustrated in Fig.~\ref{fig:FSE} in the case of the correction $\delta R_\pi$ to the decay $\pi^+ \to \mu^+ \nu [\gamma]$ (see Eq.~(\ref{eq:RPS})), evaluated for $\Delta E_\gamma = \Delta E_\gamma^{max} = M_\pi (1 - m_\mu^2 / M_\pi^2) / 2 \simeq 30$ MeV. 
It can be seen that a residual finite-volume dependence is visible in the subtracted lattice data.
The largest residual FSE  is observed when all point-like terms up to $O(1/L^2)$ are subtracted, suggesting that the structure dependent effects at $O(1/L^2)$ are important.

\section{Results for charged pion and kaon decays}
\label{sec:results}

We now insert the various ingredients described in the previous Sections in the master formula (\ref{eq:RPS}) in the case of the decays $\pi^+ \to \mu^+ \nu [\gamma]$ and $K^+ \to \mu^+ \nu [\gamma]$.
Throughout this Section it is understood that $M_{PS}$ represents the charged PS mass $M_{PS^+}$, which includes e.m.~and strong IB corrections.
The latter are calculated as in Refs.~\cite{deDivitiis:2013xla,deDivitiis:2011eh} and are generated by the mass difference $\delta m = m_d - m_u$ previously determined.

\begin{figure}[htb!]
\begin{center}
\includegraphics[scale=0.75]{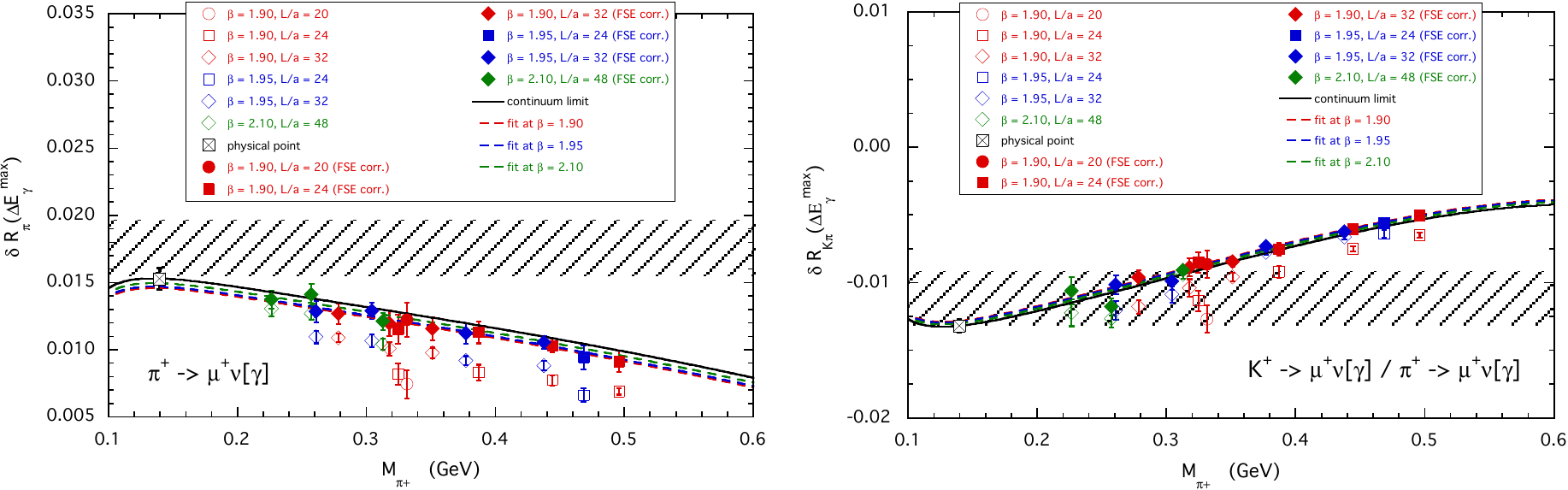}
\end{center}
\vspace{-0.75cm}
\caption{\it \footnotesize Results for the corrections $\delta R_\pi$ (left panel) and $\delta R_{K \pi} \equiv \delta R_K - \delta R_\pi$ (right panel) obtained after the subtraction of the ``universal'' FSE terms of Eq.~(\protect\ref{eq:APS_pt}) (open markers). The full markers correspond to the lattice data corrected by the residual FSEs obtained in the case of the fitting function (\protect\ref{eq:RPS_fit}) including the chiral log. The dashed lines represent the results in the infinite volume limit at each value of the lattice spacing, while the solid lines are the results in the continuum limit. The crosses represent the values $\delta R_\pi^{phys}$ and $\delta R_{K \pi}^{phys}$ at the physical point. The shaded areas correspond respectively to the values $0.0176(21)$ and $-0.0112(21)$ at 1-sigma level, obtained using ChPT (see Refs.~\protect\cite{Rosner:2015wva,Cirigliano:2011tm}).}
\label{fig:PSplus}
\end{figure}

Preliminary results for the corrections $\delta R_\pi$ and $\delta R_{K \pi} \equiv \delta R_K - \delta R_\pi$ are shown in Fig.~\ref{fig:PSplus}, where all photon energies (i.e.~$\Delta E_\gamma = \Delta E_\gamma^{max} = M_{PS} (1 - m_\mu^2 / M_{PS}^2) / 2$) are included, since the experimental data on $\pi_{\ell 2}$ and $K_{\ell 2}$ decays are fully inclusive.
The ``universal'' FSEs up to order $O(1/L)$ are subtracted from the lattice data and the combined chiral, continuum and infinite volume extrapolations are performed using the fitting function
 \be
     \delta R_{PS} = A_0 + \frac{3}{16 \pi^2} \mbox{log}(\xi) + A_1 \xi + A_2 \xi^2 + D a^2 + \frac{K_2}{M_{PS}^2 L^2} + \frac{K_2^\ell}{E_\ell^2 L^2} +
                              \delta \Gamma^{pt}(\Delta E_\gamma^{(max)})~ ,
     \label{eq:RPS_fit}
 \ee
where $\xi \equiv M_{PS}^2 / (4 \pi f_0)^2$, $E_\ell$ is the lepton energy in the meson rest frame, $A_{0, 1, 2}$, $D$, $K_2$ and $K_2^\ell$ are free parameters, and the coefficient of the chiral log is taken from Ref.~\cite{Knecht:1999ag}.

Adopting different fitting functions and FSE subtractions in order to estimate systematic errors and averaging the corresponding results, we finally get at the physical point 
 \bea
      \label{eq:Rpi_phys}
      \delta R_\pi^{phys} & = & + 0.0169 ~ (8)_{stat + fit} ~ (11)_{chiral} ~ (7)_{FSE} ~ (2)_{a^2} = + 0.0169 ~ (15) ~ , \\
      \label{eq:RKpi_phys}
      \delta R_{K \pi}^{phys} & = & - 0.0137 ~ (11)_{stat + fit} ~ (6)_{chiral} ~ (1)_{FSE} ~ (1)_{a^2} = - 0.0137 ~ (13) ~ ,
 \eea
where the errors do not include the QED quenching effects.
Our findings (\ref{eq:Rpi_phys})-(\ref{eq:RKpi_phys}) can be compared with the corresponding ChPT predictions $0.0176(21)$ and $-0.0112(21)$ \cite{Rosner:2015wva,Cirigliano:2011tm}.

\section*{Acknowledgments}
\footnotesize{We gratefully acknowledge the CPU time provided by PRACE under the project Pra10-2693 and by CINECA under the specific initiative INFN-LQCD123 on the BG/Q system Fermi at CINECA (Italy).}

\end{document}